# An external validation of Thais' cardiovascular 10-year risk assessment in the southern Thailand

Suthara Aramcharoen[1], MD; Ponlapat Satian[2], MD; Ponlachart Chotikarn[3], PhD; Sipat Triukose[4,5], PhD

[1] Thungsong Hospital, Nakhonsrithammarat, Thailand

[2] Lansaka Hospital, Nakhonsrithammarat, Thailand

[3] Marine and Coastal Resources Institute, Prince of Songkla University, Thailand

[4] Chulalongkorn University Big Data Analytics and IoT Center (CUBIC), Chulalongkorn University, Thailand

[5] Research group on Applied Computer Engineering Technology for Medicine and Healthcare (ATM), Chulalongkorn University, Thailand

## Abstract

Cardiovascular diseases (CVDs) is a number one cause of death globally. WHO estimated that CVD is a cause of 17.9 million deaths (or 31% of all global deaths) in 2016. It may seem surprising, CVDs can be easily prevented by altering lifestyle to avoid risk factors. The only requirement needed is to know your risk prior. Thai CV Risk score is a trustworthy tool to forecast risk of having cardiovascular event in the future for Thais. This study is an external validation of the Thai CV risk score. We aim to answer two key questions. Firstly, Can Thai CV Risk score developed using dataset of people from central and north western parts of Thailand is applicable to people from other parts of the country? Secondly, Can Thai CV Risk score developed for general public works for hospital's patients who tend to have higher risk? We answer these two questions using a dataset of 1,025 patients (319 males, 35-70 years old) from Lansaka Hospital in the southern Thailand. In brief, we find that the Thai CV risk score works for southern Thais population including patients in the hospital. It generally works well for low CV risk group. However, the score tends to overestimate moderate and high risks. Fortunately, this poses no serious concern for general public as it only makes people be more careful about their lifestyle. The doctor should be careful when using the score with other factors to make treatment decision.

## Introduction

Cardiovascular diseases (CVDs) are the number one cause of death globally. In 2016, WHO estimated that 17.9 million people died from CVDs (or 31% of all global deaths). The deaths from CVDs are from individual or combined effects of high blood pressure, serum cholesterol, smoking, high blood glucose and high body mass index as shown in figure 1 [1]. Most CVDs can be prevented by modifying lifestyle to avoid behavioral risk factors or by preventive medical treatment. Therefore, it is essential to have a model that can predict cardiovascular (CV) risk of each individual. The world-first CV risk model, Framingham Risk Score, was developed based on findings from Framingham Heart Study [2].

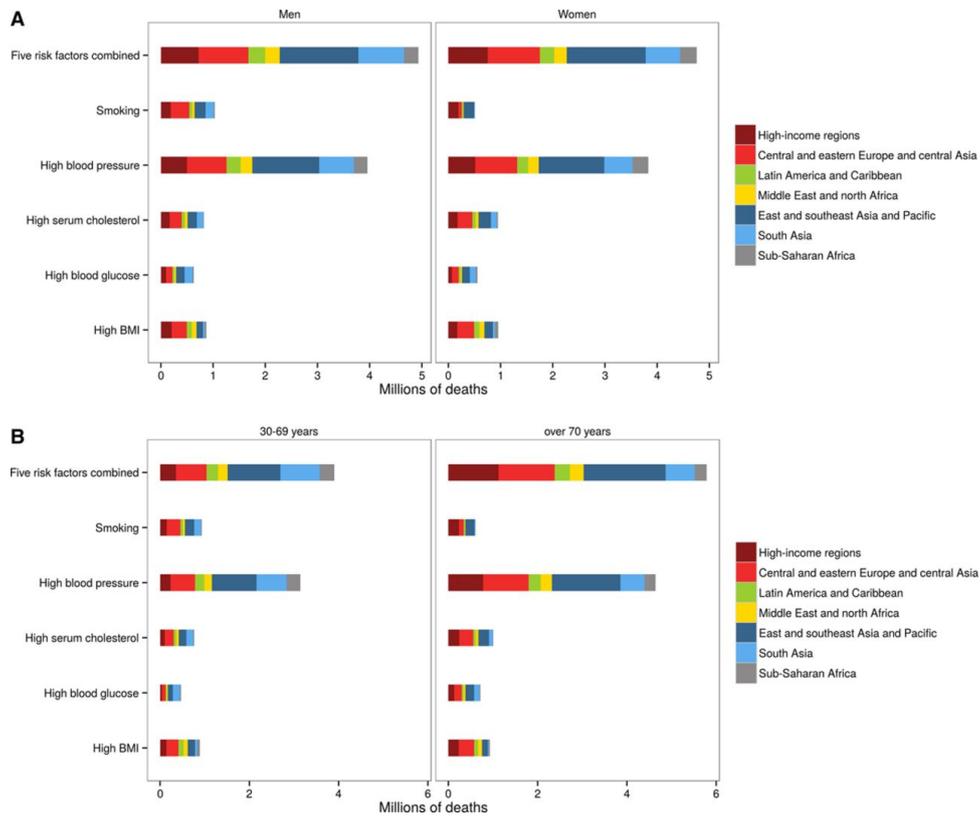

*Figure 1: The deaths from CVDs in people aged >30 years attributable to the individual or combined effects of risk factors (high blood pressure, serum cholesterol, smoking, high blood glucose and high body mass index) by region and sex (A) and age group (B) from 1980 to 2010*

In Thailand, Thai CV Risk Score [3] was developed based on findings from The Electricity Generating Authority of Thailand (EGAT) study [4], shown in the Figure 2. The model was derived from the 1st EGAT cohort comprising survey and blood sample from 3,499 EGAT's employees in Bangkok area in year 1985. The model was later on adjusted and validated using 2nd and 3rd EGAT cohorts comprising 2,999 and 2,584 samples from north western and central areas of Thailand respectively.

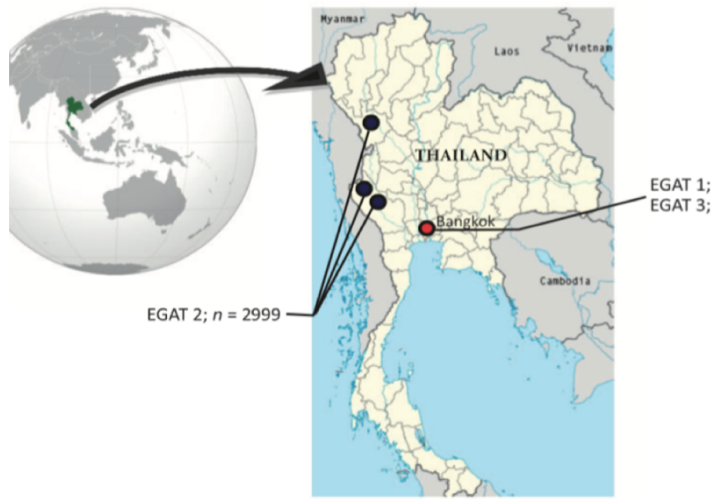

*Figure 2: Electricity Generating Authority of Thailand (EGAT) Study*

*This work is an external validation of Thai CV risk score aiming to address three issues with cardiovascular diseases as follows. Firstly, the current Thai CV Risk Score was developed using cohort of people from central and north western parts of Thailand. The question is how well the score predicts the CV risk for people from other parts of the country? Secondly, CV risk is currently being used as a guideline to prescribe drugs in statins group for patients in hospitals. A question remains unanswered that how well the score developed for general public works for hospital's patients who typically have higher risk.*

*Finally, in the age of information, it is quite convenient to gather a large set of patients' medical records, such as records from a hospital database. Medical community in UK exploits big data by using a dataset of roughly 10 million people in the development of their recent CV Risk Score, QRISK3 [5]. However, characteristics of this large dataset [8] are in many ways different from traditional cohort. Therefore, different methods to use the big data in medical research need to be studied and developed. This work is considered our first step in an attempt to use big data in the development of CV Risk Score in Thailand.*

## Methods

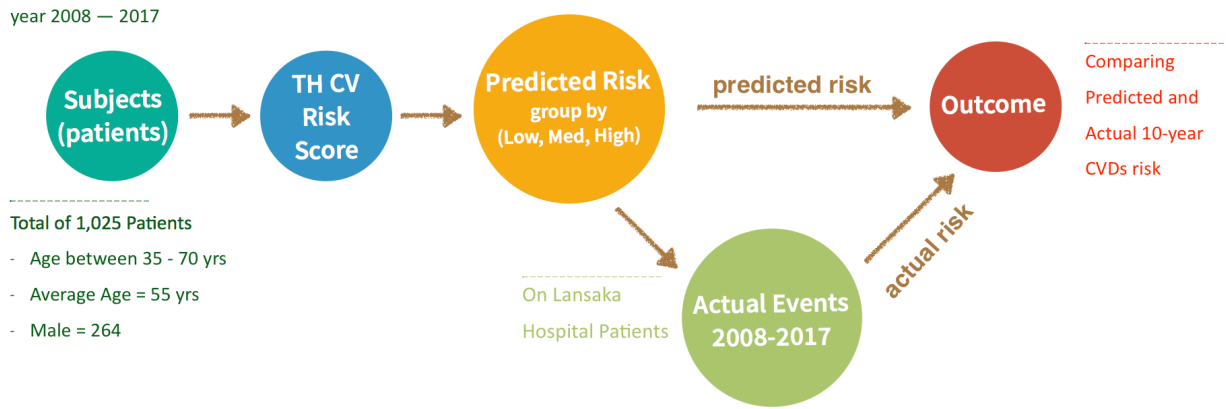

*Figure 3: Methodology used in this study*

*This work studied the use of the Thai CV Risk Score on patients at Lansaka hospital. Figure 3 illustrates methodology used in this study. The dataset contains records of 1,025 patients (319 males, 35-70 years old) with no cardiovascular event prior to 2008. We calculated 10-years cardiovascular risk using the following factors – sex, age, smoke, diabetes, systolic blood pressure, LHL and HDL -- of all patients in 2008 and observed their actual cardiovascular events from 2008 - 2017. In this study, target cardiovascular events are defined as a composite of fatal and non-fatal myocardial infarction or stroke as shown in the Table 1.*

*Table 1: Cardiovascular Events*

| Hazard Event | Subtype | ICD10 | ICD10 Group |
|---|---|---|---|
| Coronary Revascularization | - | Z95 | Z95.5 |
| Angina Pactoris | - | I209 | - |
| Unstable Angina | - | I200 | I200-I209 |
| Myocardial Infarction | NSTEMI | I214 | I210-I219 |
| | STEMI | I213 | |
| CHD Death | - | I519 | I510-I519 |
| Stroke | Hemorrhagic Stroke | I619 | I600-I64 |
| | Ischemic stroke | I64 | |
| Cardiac Failure | - | I509 | I500-I509 |

## Results

Table 2 presents validation outcome of Thai CV risk score. The risk score indicates chances of having cardiovascular events in 10-year time. The score below 0.1, between 0.1 and 0.2, and more than 0.2 are interpreted as low, moderate, and high risks respectively. Therefore, we grouped Lansaka's patients into 3 groups based on their predicted risk level and calculate actual risk for each group from cardiovascular events found between 2008 and 2017. In low risk group, on average, the risk score slightly underestimated the actual risk by 0.0068. However, in both moderate and high-risk groups, the score, on average, overestimated the actual risk by 0.037 and 0.088 respectively.

*Table 2: The validation outcome of Thai CV risk score*

| | Actual | Predicted (group average) | Estimation |
|---|---|---|---|
| Low Risk [0,10%) | 0.0707 | 0.0529 | As Expected |
| Moderate Risk [10%,20%) | 0.0884 | 0.1522 | Over Estimate |
| High Risk [20%,30%) | 0.1364 | 0.2523 | Over Estimate |

## Conclusion

In brief, Thai CV risk score works for southern Thais population including patients in the hospital. It generally works well for low CV risk group. However, the score tends to overestimate moderate and high risks and the error goes up as the risk increases. Fortunately, this poses no serious concern for general public as it only makes people be more careful about their lifestyle. The doctor should be careful when using the score with other factors to make treatment decision.

## Limitations

The data used in this study are from Lansaka hospital's patient records. Due to several factors, such as migration, the 10-years dataset, from 2008 – 2017, might be incomplete. For example, if the patient moved to another city after 2008 and having cardiovascular event later on, the Lansaka's dataset will indicate that this patient never have any CVD events, which is inaccurate. This is a type I censoring that leads to right censoring condition [7]. Patients who stay with Lansaka hospital for the entire 10-years period but have incomplete information for assessing CV risk are also excluded from our study. We make it our future work to resolve all of these limitations.

## Future Work

A set of risk factors used in the current Thai CV risk score is similar to the original Framingham risk score [2]. However, the more recent CV risk score, such as QRISK3 [5], uses more extensive set of factors. Our research team aims to extend the current Thai CV risk score to includes additional factors that can help improve the assessment of the CV risk for all Thai.

The data required to calculate the 10-year risk score are available for a small fraction of patients at Lansaka hospital. It is mainly due to missing cholesterol values, which is understandable since the blood draw is required to acquire this data. In this case, the waist to height ratio is used in place of cholesterol to assess CV risk. We are exploring a non-invasive approach to obtain these missing values and currently looking into using deep learning on retinal fundus images [6] to infer the missing information.